\newcommand{\udl}{U}
\documentclass[aps,preprint,english,prb,superscriptaddress]{revtex4}
\usepackage[T1]{fontenc}
\usepackage[latin1]{inputenc}
\usepackage{float}
\usepackage{graphicx}
\usepackage{pxfonts}
\usepackage{verbatim} 
\usepackage{amssymb}

\makeatletter

\usepackage{graphicx}
\usepackage{babel}
\usepackage{setspace}

\makeatother
\begin{document}
\title{Weighted Ensemble Path Sampling for Multiple Reaction Channels}
\author{Bin W. Zhang}
\affiliation{Department of Computational Biology, School of Medicine, University of Pittsburgh, Pennsylvania 15213}
\author{David Jasnow}
\affiliation{Department of Physics \& Astronomy, University of Pittsburgh, Pittsburgh, Pennsylvania 15260}
\author{Daniel M. Zuckerman \footnote{Electronic mail: ddmmzz@pitt.edu}}
\affiliation{Department of Computational Biology, School of Medicine, University of Pittsburgh, Pennsylvania 15213}
\date{\today}
\begin{abstract}
Finding and sampling multiple reaction 
channels for molecular transitions remains an important challenge in physical chemistry. Here we show that the weighted ensemble (WE) path sampling method can readily sample multiple channels. 
In a first test, both the WE and transition 
path sampling methods are applied to two-dimensional model potentials. The comparison explains why the
weighted ensemble approach will not be trapped in one channel. 
The WE approach is then used to sample the full transition path ensemble in implicitly solvated alanine dipeptide at two different temperatures.
The ensembles are of sufficient quality to permit quantification of the fractional importance of each channel, even at $T=300K$ when brute-force simulation is prohibitively expensive.
\end{abstract}

\maketitle

\section{INTRODUCTION}
Path sampling is an important strategy for studying study conformational transitions and activated processes.
Some path sampling methods were applied to large biological system recently, for example, to
study the conformational transitions of proteins. 
Conformational transition is important for protein to carry out their functions. But the
pathway of these transitions are not easy to obtain either by experiments or simulations.
Experiments, such as resolution studies by X-ray or NMR, are capable to determine the atomic
structure of the stable states of protein, but not the transition states in the middle of the
pathway. Conventional MD simulations are not good to study activated process because it
involves a waiting step, which has been studied for a long time. The long waiting time to generate an
ensemble of activated processes by straightforward brute-force approach is usually beyond the
reach of today's computational power. In fact, if the activated process does take place, it is
usually very quickly. Path sampling approaches take advantage of this property by focusing 
computer resources exclusively on rare transition events.

A number of path generating and path sampling approaches have been developed. 
Some \emph{ad hoc} path generating methods have been applied for large biological systems, like the targeted 
molecular dynamics (TMD) 
~\cite{Schlitter-TMD_1-MS-1993,
Schlitter-TMD_2-JMG-1994,
Karplus_Vaart-Restricted_Perturbation_TMD-JCP-2005,
Karplus_Vaart-Minimum_Free_Energy_Pathways-JCP-2007}
and steered molecular dynamics (SMD).
~\cite{Schulten_Izrailev-SMD_Avidin_Biotin_Complex-BJ-1997,
Schulten_Lu-Titin_SMD-BJ-1998,
Schulten_Marszalek-SMD_Titin-Nature-1999}
Another type of path sampling approaches focus on the optimal paths or the paths close to the optimal paths,
such as the milestone method and the string method.
Some path sampling methods take root in the idea of path integral methods, they sample the paths of rare events 
by evaluating their relative weights, such as, the transition path sampling (TPS),
~\cite{Chandler-Transition_Path_Sampling_Application-JCP-1998,
Chandler-Transition_Path_Sampling_Reaction_Rate-JCP-1998,
Bolhuis_Chandler-Transition_Path_Sampling-ARPC-2002} 
the dynamic importance sampling (DIMS),
~\cite{Zuckerman-Important_Sampling-JCP-1999,
Zuckerman-Important_Sampling-PRE-2000,
Woolf_Jang-Multiple_Pathways_Alanine_Dipeptide-JCC-2006} 
the transition interface sampling approach (TIS)
~\cite{VanErp_Bolhuis-Path_Sampling_Rate_Constant-JCP-2003,
VanErp_Bolhuis-Elaborating_transition_interface_sampling_methods-JCOP-2005} 
and the random walk in path space by following a 
pseudo Langevin equation. In principle those methods based on the calculation of the weights of paths 
can generate properly distributed path ensembles. 
Not long ago Zuckerman, Jasnow and Zhang showed the Weighted Ensemble (WE) method
~\cite{Huber_Kim-WEB-BJ-1996,
Subramaniam-WEB_Protein-PNAS-1998,
Zhang_Jasnow_Zuckerman-WE_CaM-PNAS-2007}
is a very promising simulation approach to investigate conformational transitions, which can lead to 
correct path ensembles and reaction rate simultaneously.
WE is also an exact algorithm for many dynamics types.
~\cite{Zuckerman_Jasnow-Arxiv}

Despite a number of successes, path sampling research still faces many challenges.
One key difficulty is finding multiple reaction channels for a transition.
~\cite{Bolhuis-Rare_Events_Multiple_Channels_TPS_RE-JCP-2008}
For example, the transition path sampling approach, perhaps the best known method, can become trapped in a local 
minimum, and miss other channels. 
~\cite{Bolhuis-Rare_Events_Multiple_Channels_TPS_RE-JCP-2008}
The transition interface sampling method has the same drawback as its ancestor, TPS.
Recently Bolhuis and van Erp tried to solve this problem by using the combination of the replica exchange and the transition 
interface sampling methods (RETIS).
~\cite{Bolhuis-Rare_Events_Multiple_Channels_TPS_RE-JCP-2008,
vanErp-TIS_RE_Multiple_Reaction_Channels-CPC-2008} 

The aim of this paper is to show the weighted ensemble method has no difficulty to sample multiple transition channels.
The weighted ensemble method, which is based on unbiased replication and combination of brute-force simulations,
does not have this drawback.

This paper is set up as follows. In Sec.\ref{Sec:method}, we introduce the models and review the methods briefly. In Sec.\ref{Sec:result},
first the weighted ensemble and transition path sampling methods are applied to two-dimensional model potentials. This simple example
highlights the differences between the two methods sampling of multiple channels. Then the weighted ensemble approach is tested to find
the transition events between two stable structures of alanine dipeptide at high temperature $500K$. The path ensembles are compared 
with those gained from brute-force simulations. At last the weighted ensemble method samples the \emph{full} transition ensemble of alanine dipeptide at 
room temperature $300K$, which is prohibitively expensive for brute-force simulation. This paper ends with conclusions
and discussion.

\section{Models and Methods}\label{Sec:method}
\subsection{Two-Dimensional Models}
\begin{figure}[H]
\begin{center}
\includegraphics[width=8 cm]{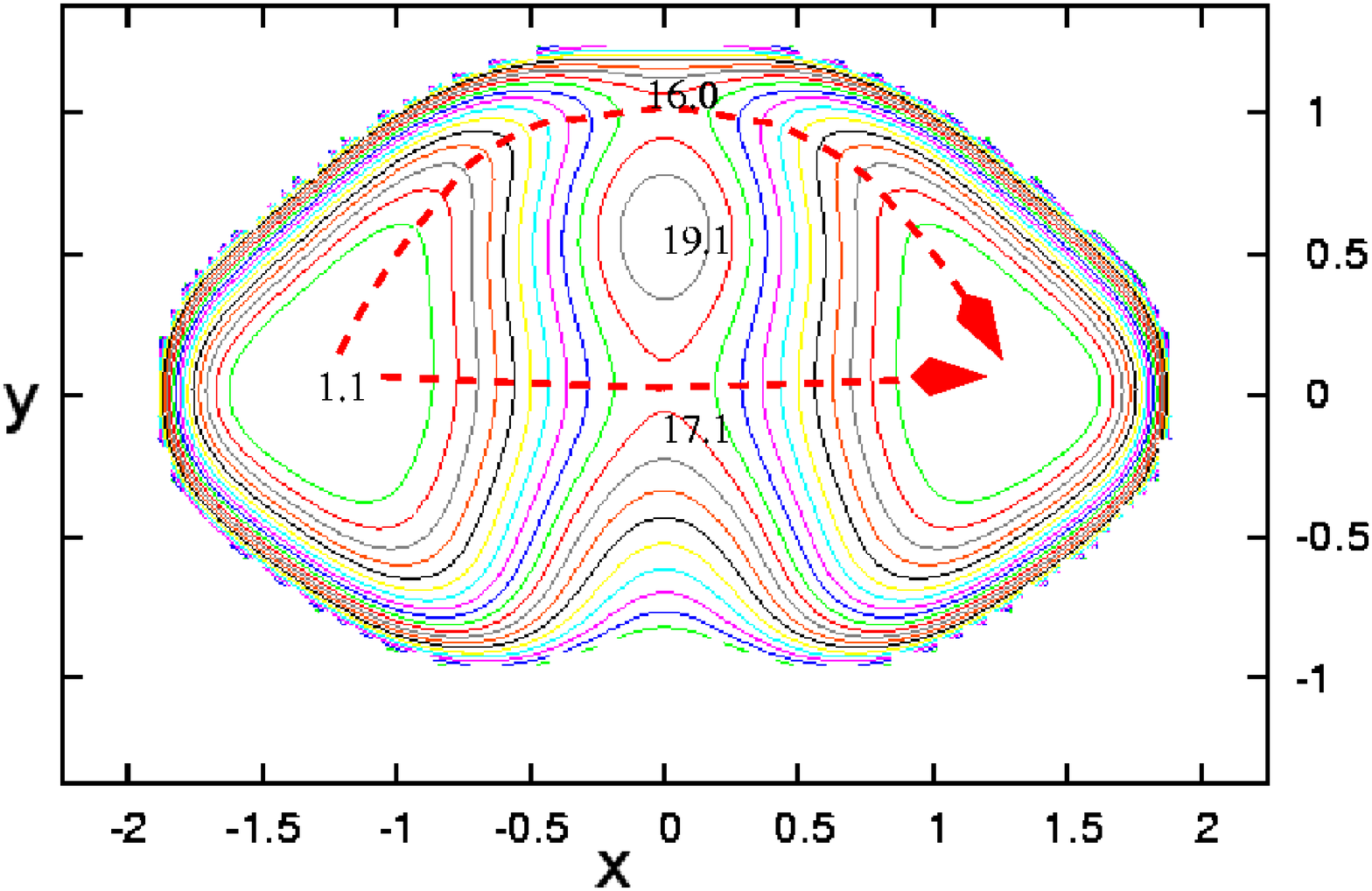}\includegraphics[width=8 cm]{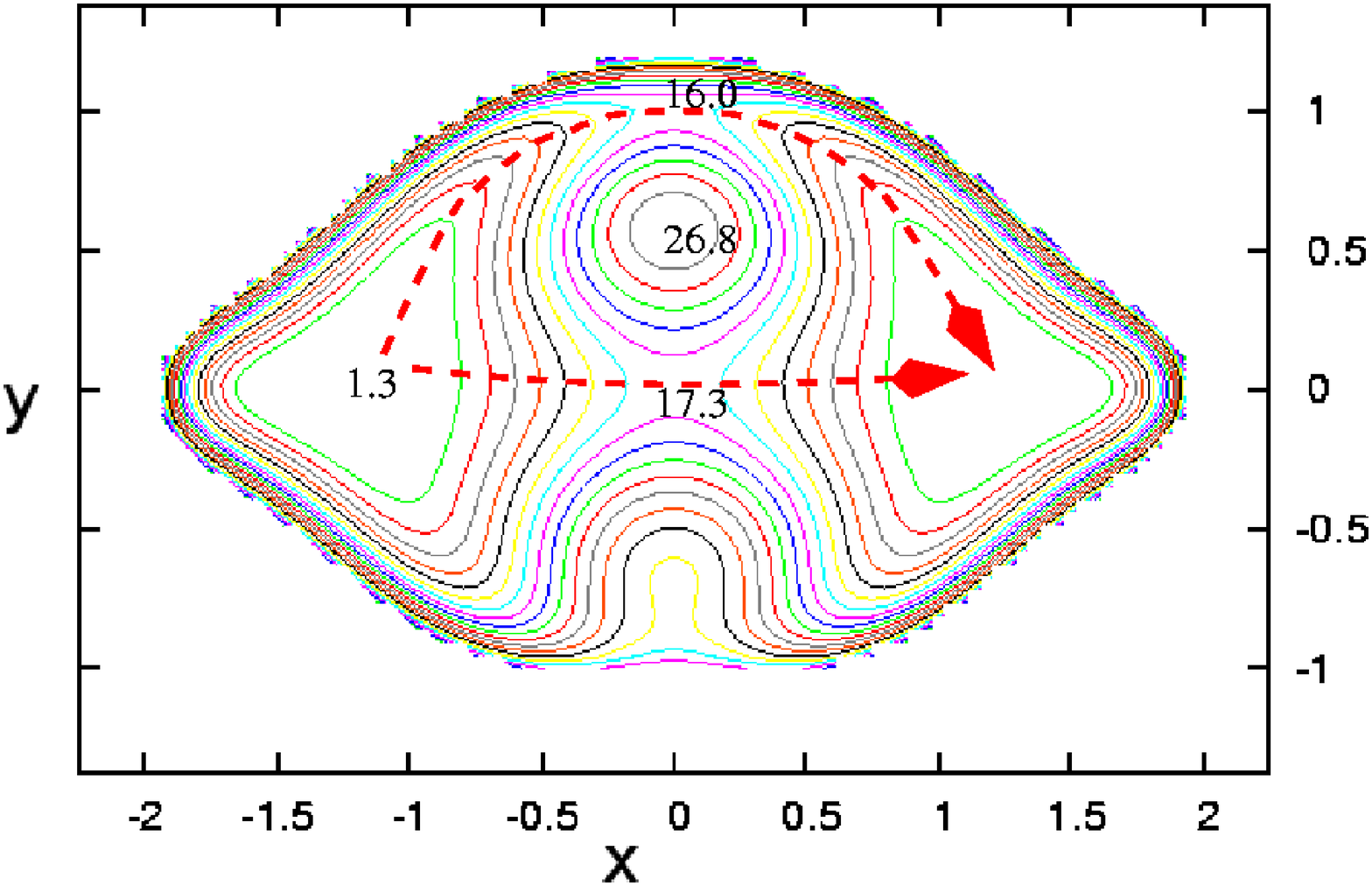}
\caption{\label{fig:Contours} Contours of two-dimensional toy potentials. The left panel is the potential $\udl_1$ with 
$\alpha=20$, the right panel is the potential $\udl_2$ with $\alpha=72$. The red arrows show the different channels
connecting the left and right wells.
The numbers among the contours indicate energy values in $k_BT$ units for extrema and saddles.
}
\end{center}
\end{figure}

We will show the difference of how the WE approach and TPS approach sample multiple channels by using two-dimensional 
model potentials. 
The toy potentials $\udl_1$ and $\udl_2$ are inspired by Chen, Nash and Horing's work,
and defined via
\begin{eqnarray}
\left. \udl(x,y) \right/ k_B T &=& \alpha(x^2+y^2-1)^2y^2\nonumber\\
&\ & -\exp\{-4[(x-1)^2+y^2]\}-\exp\{-4[(x+1)^2+y^2]\}\nonumber\\
&\ & +\exp[8(x-1.5)]+4\exp[-8(x+1.5)]\nonumber\\
&\ & +\exp[-4(y+0.25)]+16\exp(-2x^2)\,,
\label{2DT-1}
\end{eqnarray}
with $\alpha=20$ and $\alpha=72$ respectively distinguishing $\udl_1$ and $\udl_2$.
The contours of these two potentials are shown in Fig.~\ref{fig:Contours}. They both have two channels connecting
the left and right wells, which are shown by the red arrows.
The only significant difference between these two potentials is the height of barrier separating two channels. Compared with the 
saddle points ($\sim 16k_BT$ and $17k_BT$ in Fig.~\ref{fig:Contours}),
when $\alpha=20$, the barrier height is about $3k_BT$. And when $\alpha=72$, the barrier height is about $10k_BT$.
The WE approach (using the horizontal position $x$ as the progress coordinate)
and TPS approach are applied to study the transition events of an over-damped Brownian particle from the left well
to the right well. The initial state and the final state are defined as the regions where the potential satisfies $\udl(x,y)<(\udl_{min}+2)$,
where $\udl_{min}$ is the lowest potential in the left and right wells.

\subsection{Alanine Dipeptide}
\begin{figure}[H]
\begin{center}
\includegraphics[width=8 cm]{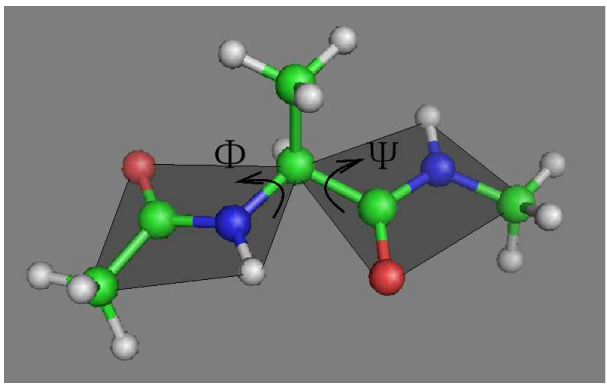}
\caption{\label{fig:AlaD} Alanine dipeptide molecule ($\mathrm{CH_3CO-Ala-NHCH_3}$). 
Different atoms are shown by different colors. White: Hydrogen; Red: Oxygen; Green: Carbon; Blue: Nitrogen.}
\end{center}
\end{figure}

The second system on which we will test the weighted ensemble approach is alanine dipeptide (ace-ala-nme).
The molecule is shown in Fig.~\ref{fig:AlaD}. 
The principle variables describing the structure of  alanine dipeptide are two
backbone dihedral angles: $\Phi$ (C-N-C-C) and $\Psi$ (N-C-C-N). 
Alanine dipeptide is frequently used for testing simulation methods and force fields.
~\cite{Brooks_Tobias-AlaD_In_Gas_Phase_And_Aqueous_Solution-JPC-1992}
The reasons of choosing this molecule are as following.
First it is one of the simplest molecules which contains two full peptide planes,
so it will contains many structural features of protein backbones.
~\cite{Branden_Tooze-Introduction_2_Protein_Structure-1st-1991}
Second alanine dipeptide is small enough that its free energy surface can be studied thoroughly 
by different approaches.
~\cite{Brooks_Tobias-AlaD_In_Gas_Phase_And_Aqueous_Solution-JPC-1992,
Friesner_Philipp-Mixed_QM_MM_Modeling-JCC-1999,
Caflisch_Apostolakis-Conformational_Transitions_And_Barriers_Alanine_Dipeptide_In_Water-JCP-1999,
McCammon_Marrone-Comparison_Models_Of_Solvation_PMF_AlaD-JPC-1996,
Pappu_Drozdov-Solvent_Determining_Conformational_Preferences_AlaD-JACS-2004,
Karplus_Smith-Stochastic_Simulations_AlaD-JPC-1993,
Smith-Alanine_Dipeptide_Free_Energy_Surface_In_Solution-JCP-1999,
Vargas_Dixon-Conformational_Study_Of_AlaD-JPCA-2002}
Third the conformational transitions of alanine dipeptide contains multiple channels and 
have been studied by several groups recently.
~\cite{Woolf_Jang-Multiple_Pathways_Alanine_Dipeptide-JCC-2006,
Branduardi-From_A2B_In_Free_Energy_Space-JCP-2007,
Karplus_Vaart-Minimum_Free_Energy_Pathways-JCP-2007}

\begin{figure}[H]
\begin{center}
\includegraphics[width=8 cm]{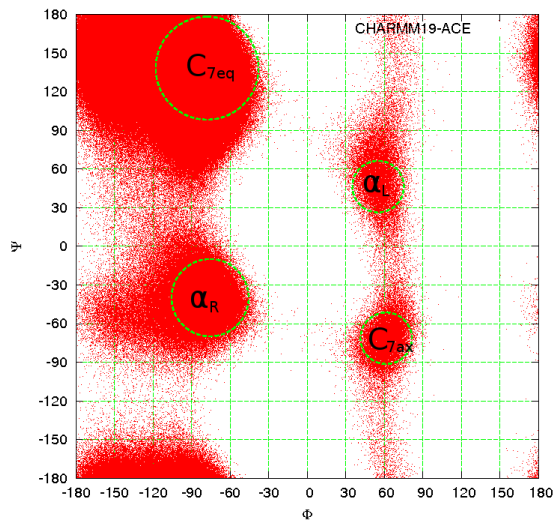}
\caption{\label{fig:AlaD-EM}Four stable states of alanine dipeptide.
The dihedral angles $\Psi$ and $\Phi$ are chosen as the progress coordinates for the WE approach.
The two-dimensional space of dihedral angles is divided into a $12\times12$ grid for WE simulation, with $20$ 
simulations allowed in each grid.}
\end{center}
\end{figure}

Several brute force simulations are performed first, using Langevin dynamics in the CHARMM program 
~\cite{Karplus_Brooks-CHARMM-JCC-1983} 
to find the energy minimum states.
The simulations use the ``united atom model'' with the CHARMM parameter set 19 and 
implicit solvent ACE (analytical continuum electrostatics) model. 
~\cite{Karplus_Schaefer-ACE-JPC-1996}
The dihedral angles of the four energy minimum states we find are shown in Table.~\ref{tb:aladmin}.
Our locations of minima are somewhat different with the previous study,
~\cite{Smith-Alanine_Dipeptide_Free_Energy_Surface_In_Solution-JCP-1999,
Caflisch_Apostolakis-Conformational_Transitions_And_Barriers_Alanine_Dipeptide_In_Water-JCP-1999,
Karplus_Vaart-Minimum_Free_Energy_Pathways-JCP-2007,
Woolf_Jang-Multiple_Pathways_Alanine_Dipeptide-JCC-2006}
but it is known that the simulation of alanine dipeptide is very sensitive to solvent model.
~\cite{Woolf-Path_Corrected_Functionals_Of_Stochastic_Trajectories-CPL-1998,
Woolf_Jang-Multiple_Pathways_Alanine_Dipeptide-JCC-2006}
The WE approach is applied to study the transition events between state $C_{7eq}$ and $C_{7ax}$.
The initial state $C_{7eq}$ is defined as the area closed by the circle 
\begin{equation}
[\Psi-(-77.9)]^2+[\Phi-(138.4)]^2=(40)^2\,,
\label{AlaD-1}
\end{equation}
and the final state $C_{7ax}$ is the area closed the circle
\begin{equation}
[\Psi-(61.4)]^2+[\Phi-(-71.4)]^2=(20)^2\,,
\label{AlaD-2}
\end{equation}
as shown in Fig.~\ref{fig:AlaD-EM}.

We choose the dihedral angles $\Psi$ and $\Phi$ as the progress coordinates for the WE approach.
The two-dimensional space of dihedral angles is cut into a $12\times12$ grid, with $20$ 
simulations allowed in each grid. After every $\tau=100\mathrm{fs}$, the embedded CHARMM simulations 
are paused, and the simulations are combined and split without bias.
The weighted ensemble program was stopped after $2500\tau$.

\section{Results}\label{Sec:result}
\subsection{Two-Dimensional Models}
\begin{figure}[H]
\begin{center}
\includegraphics[width=6 cm]{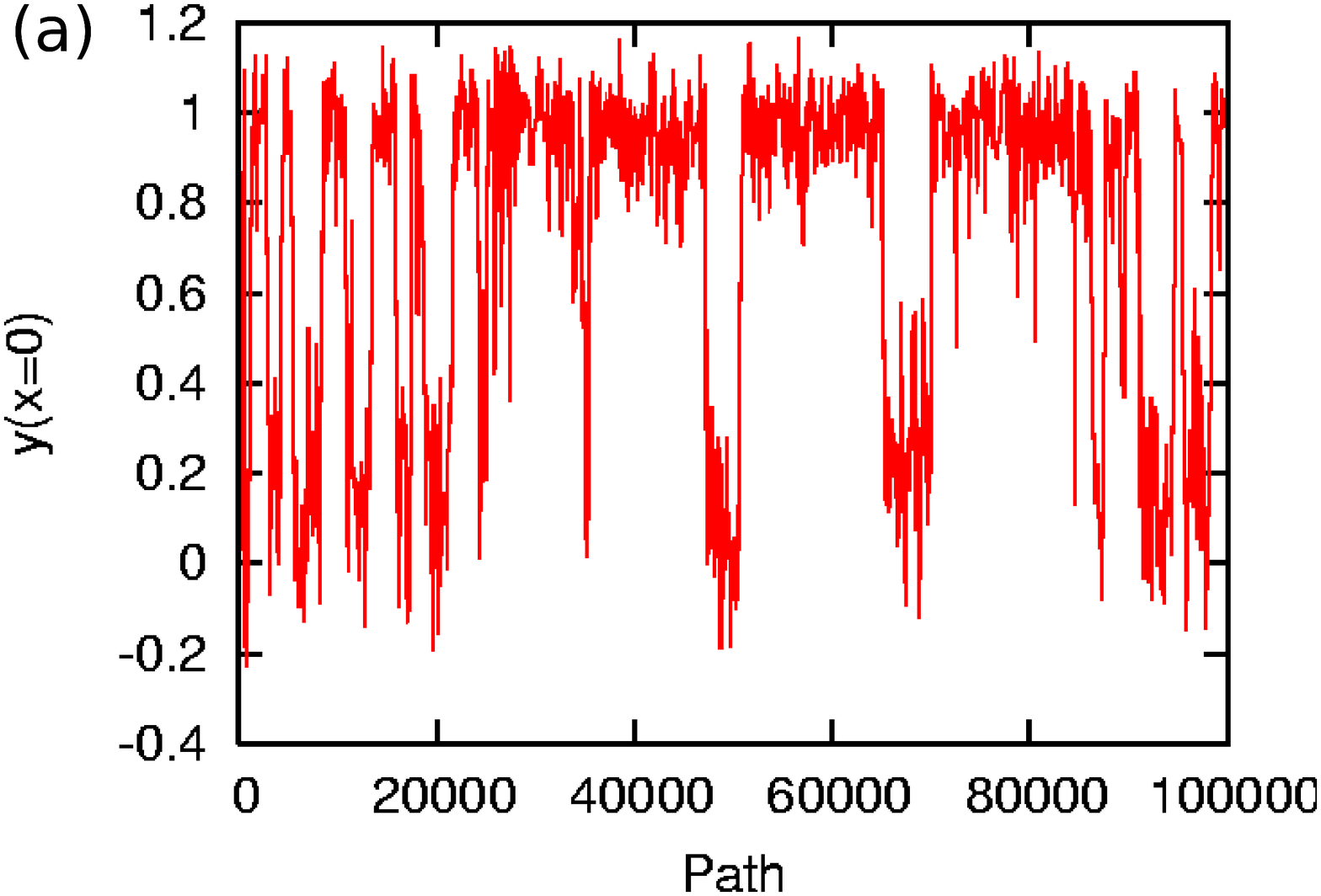}\includegraphics[width=6 cm]{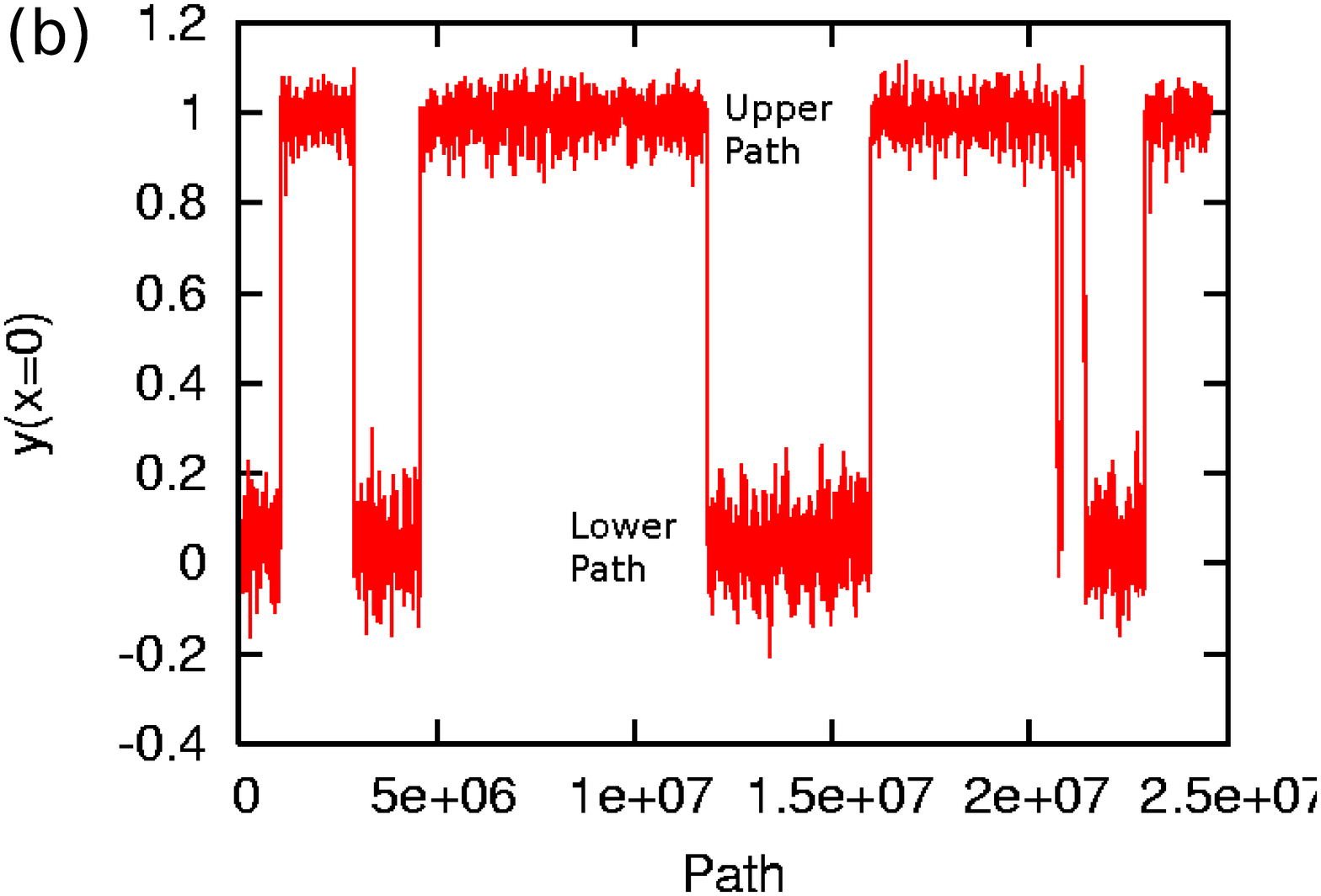}\\
\includegraphics[width=6 cm]{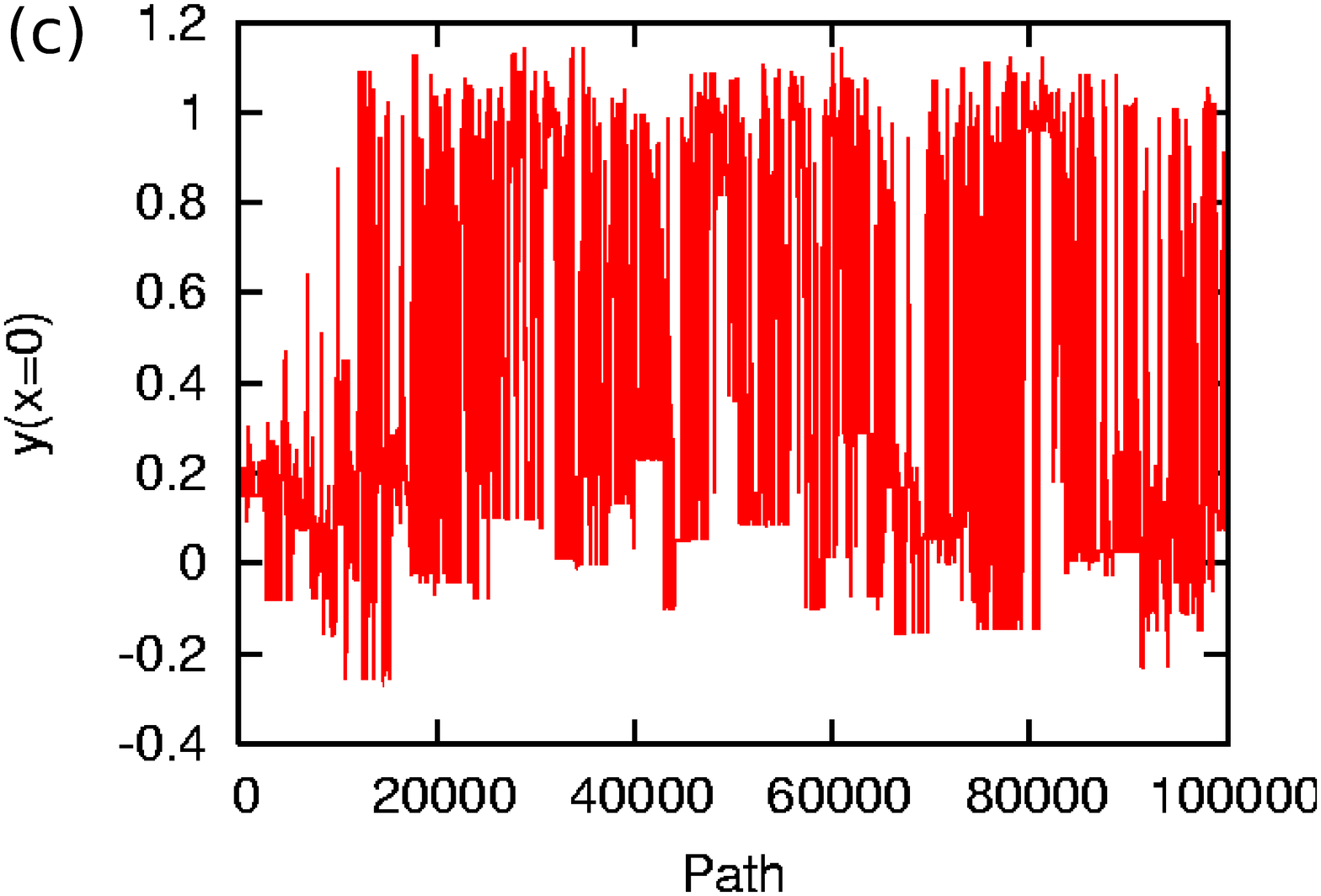}\includegraphics[width=6 cm]{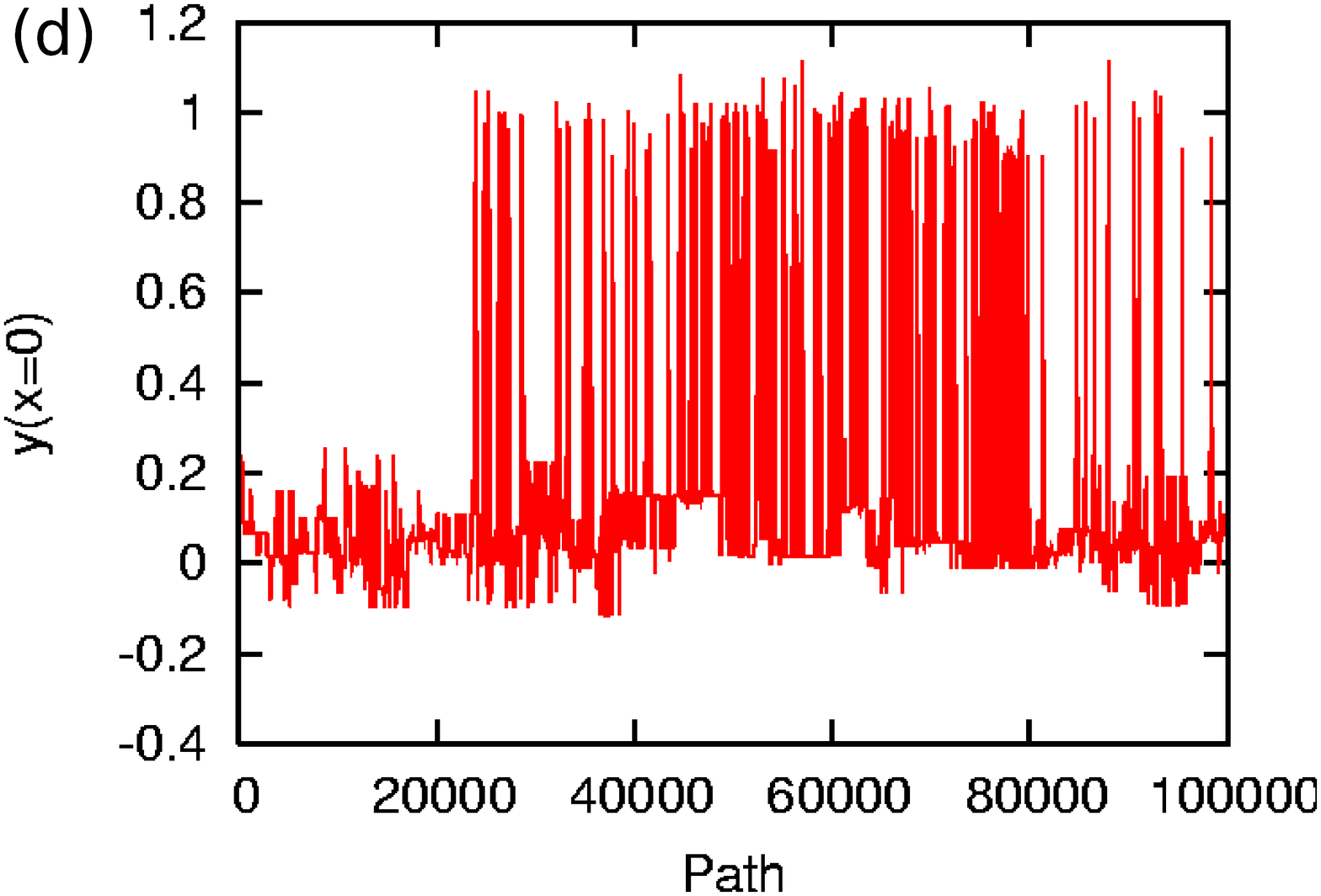}\\
\caption{\label{fig:PathSwitch}
Channel-switching in TPS and WE simulations of toy models.  Both methods were applied to the two toy potentials, $\udl_1$ and $\udl_2$ of Fig.\ \ref{fig:Contours}.  The various panels show (a)TPS for $\udl_1$, (b)TPS for $\udl_2$,
(c)WE for $\udl_1$ and (d)WE for $\udl_2$. The TPS method samples the channels one by one. 
In panel (a), for potential $\udl_1$, the  switch happens about every $10^4$ paths.
In panel (b), for potential $\udl_2$, the switch happens approximately every $5\cdot 10^6$ paths. 
But in panel (c) and (d), after a short transient period, WE simulations generate different types of 
paths simultaneously.
Note the greatly enlarged horizontal scale in panel (b).
}
\end{center}
\end{figure}

The TPS approach and the WE approach sample the paths in multiple channels in different ways. To show the difference,
we use the $y$ position of path crossing the $x=0$ section last time to identify which channel it belongs to, and plot this
position for each path versus the path index (simulation time).  Fig.~\ref{fig:PathSwitch}(a) 
and Fig.~\ref{fig:PathSwitch}(b) show the path switching of TPS method for potentials $\udl_1$ and $\udl_2$. 
The TPS method, as a Monte Carlo simulation, samples the channels (local minimum in the path space) one by one. 
For potential $\udl_1$, the  switch happens about every $10^4$ paths, 
For potential $\udl_2$, because of the higher barrier, the switch happens approximately 
every $5 \cdot 10^6$ paths, which is much less frequent. This reveals how, for higher barriers, the sampling can
get ``trapped'' and could be biased.
Fig.~\ref{fig:PathSwitch}(c) and Fig.~\ref{fig:PathSwitch}(d) show the path switching of WE method for these two
potentials. The frequent switches suggest that the Weighted Ensemble method, after a short transient period,
generates different types of paths simultaneously in both potentials. 

\begin{figure}[H]
\begin{center}
\includegraphics[width=6 cm]{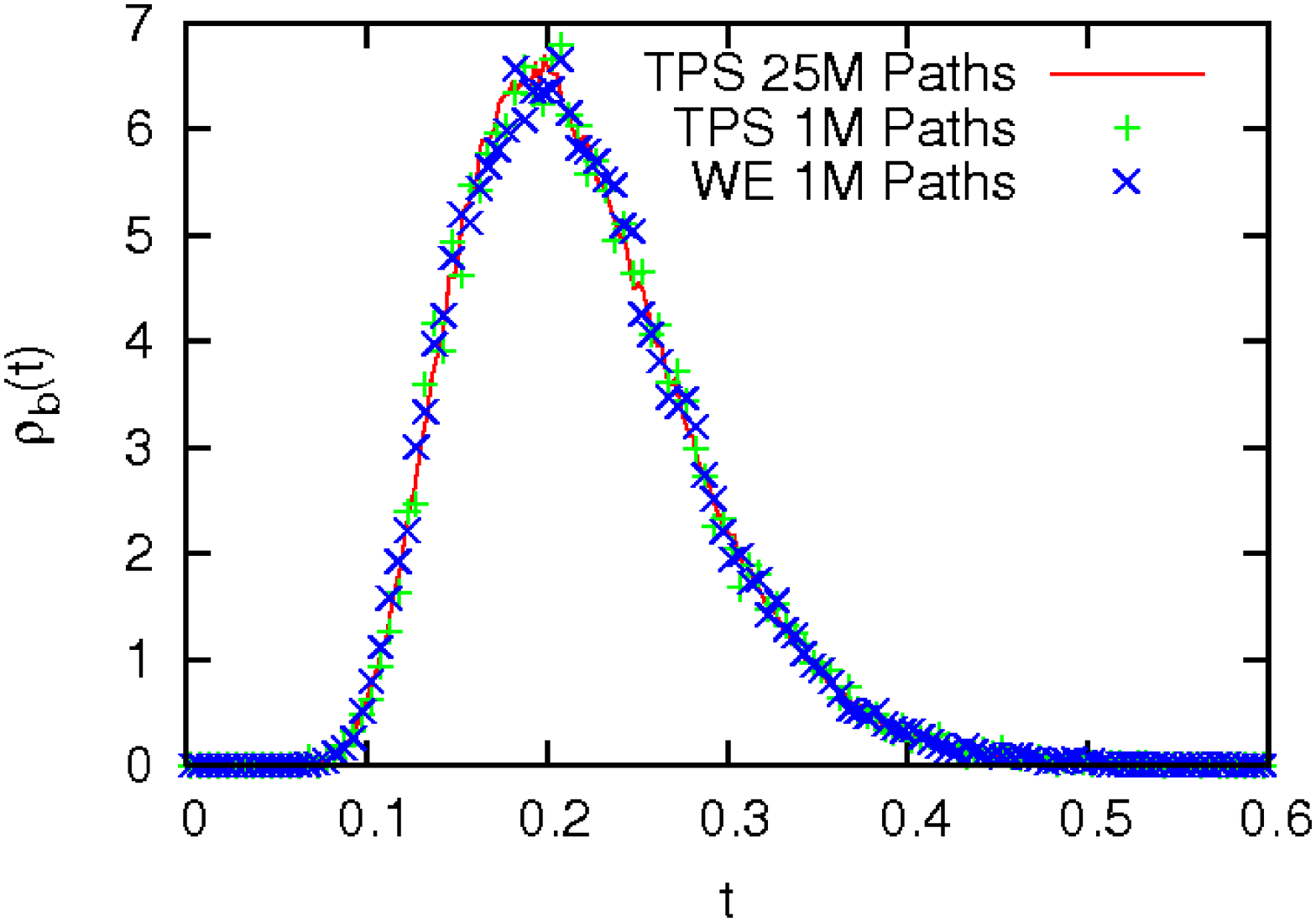}\includegraphics[width=6 cm]{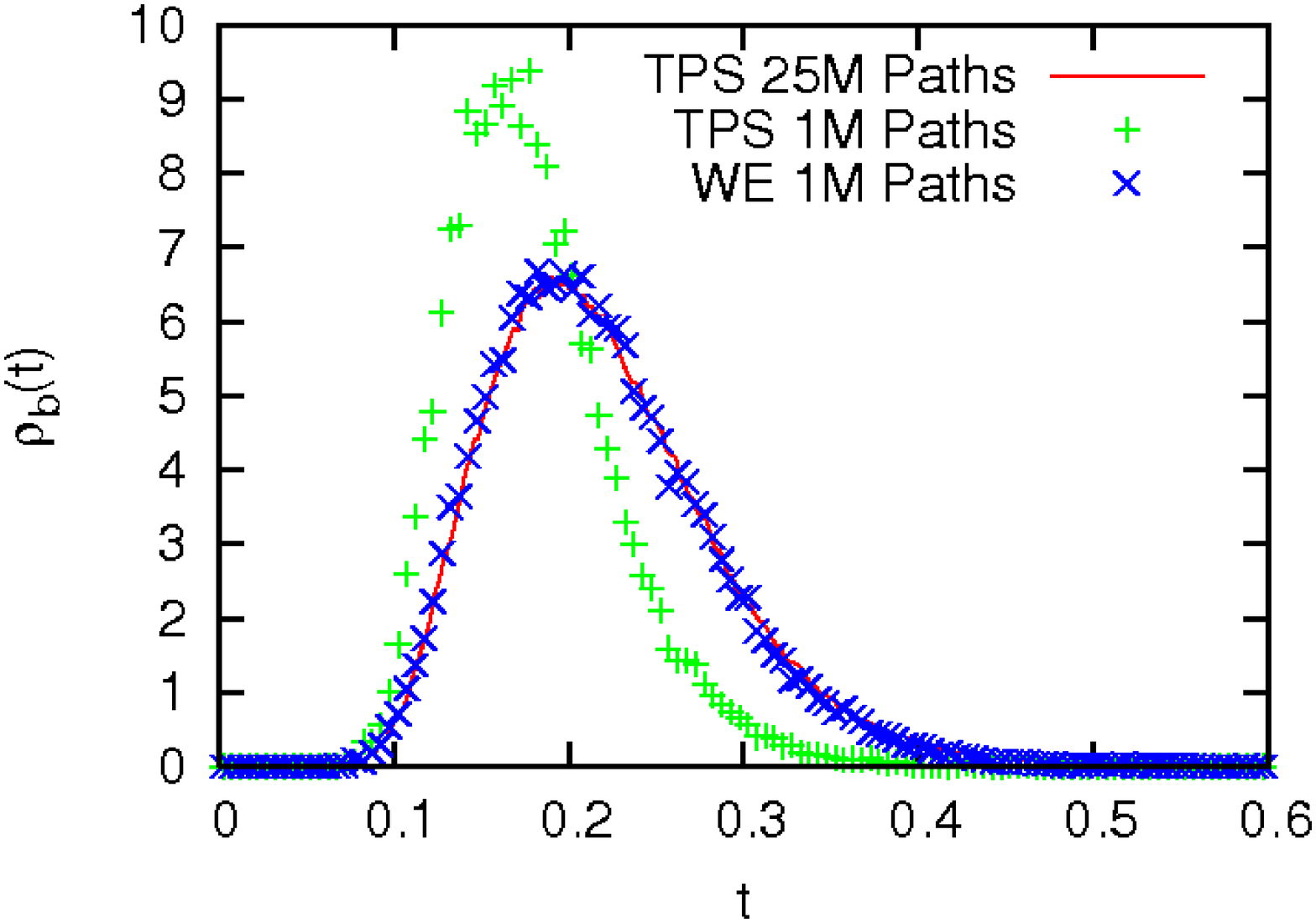}
\caption{\label{fig:Tb2D}Distribution of transition-event durations in toy models. The left panel is
for potential $\udl_1$ (smaller barrier), the right panel is for potential $\udl_2$ (larger barrier).
Notice in the right panel, a ``short'' transition path sampling simulation yields an incorrect 
distribution.}
\end{center}
\end{figure}

Both methods are used to get the distribution of transition-event durations. The definition of the transition-event duration
is the time interval between the last time the Brownian particle leave the initial state and the first time it reaches the
final state. It is the time the Brownian particle uses to finish the transition. The WE simulations are stopped after they 
generate $10^6$ paths. The TPS simulations are stopped after $2.5\cdot 10^7$ paths are obtained. 
The results are shown in Fig.~\ref{fig:Tb2D}. For small barrier ($\udl_1$), both methods get the correct distribution
by using $10^6$ paths. 
But for potential $\udl_2$, because of the high barrier separating the two types of paths, a ``short'' 
transition path sampling simulation yields an incorrect 
distribution after $10^6$ paths have been sampled. In that case, the transition path sampling is 
trapped by the local minimum in path space; see Fig.~\ref{fig:PathSwitch}(b).

\subsection{Alanine Dipeptide}
The WE method is applied to study the transition events between state $C_{7eq}$ and $C_{7ax}$ of alanine dipeptide under 
temperature $300K$ and $500K$. The brute-force simulations are also used to get the transition paths under $500k$, and the
results are compared with those given by WE method. 

\subsubsection{The Transition Rate}
The reaction rate $k$ is an important quantity for chemical and biological reactions and transitions. If the first passage
time is too long, this quantity will be impossible to obtain by brute-force simulations. The WE method 
can yield the path ensemble and reaction rate simultaneously. Under $500K$ brute-force simulations obtained the reaction
rate $k_{\mathrm{BF}}=1.5 \times 10^{-1}/ns$. And WE method obtained the reaction rate $k_{\mathrm{WE}}=1.4 \times 10^{-1}/ns$.
They are in good agreement. When temperature is $300K$, the reaction rate obtained by WE method is $k_{\mathrm{WE}}=1.6 \times 10^{-3}/ns$,
which means compared with $500K$, the transition is about $100$ times more difficult to happen now.

\subsubsection{Paths in Different Channels}
\begin{figure}[H]
\begin{center}
\includegraphics[width=8 cm]{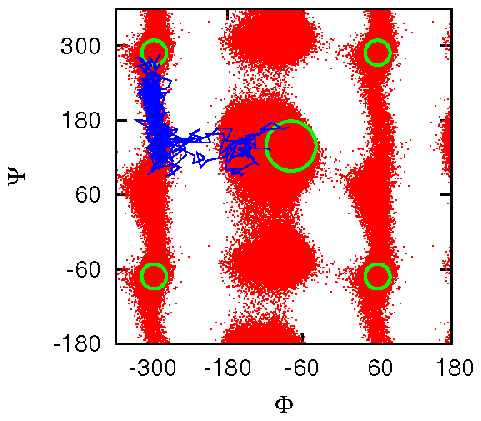}\includegraphics[width=8 cm]{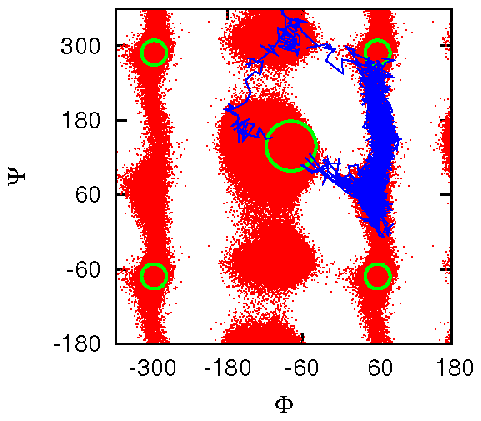}\\
\includegraphics[width=8 cm]{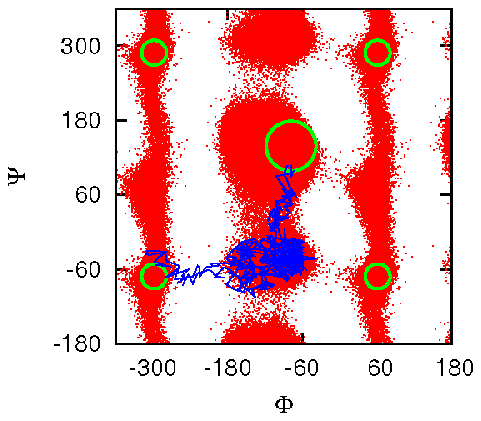}\includegraphics[width=8 cm]{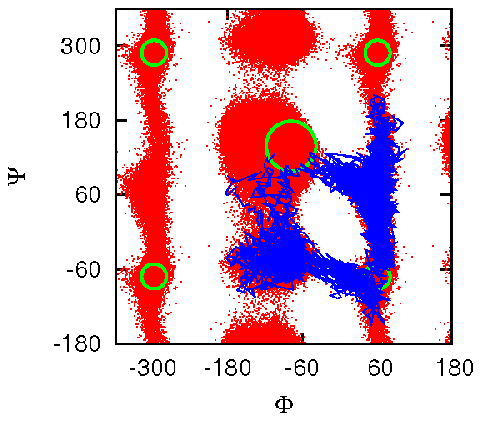}
\caption{\label{fig:AlaD-4Ch} Four types of transition paths between the states $C_{7eq}$ and
$C_{7ax}$ of alanine dipeptide at $300K$.  Approximately $100$ paths are randomly chosen based on their weights (resampled) from the full path ensemble.
Different paths traverse different barriers, and therefore belong to different channels.}
\end{center}
\end{figure}

The path ensemble will be studied in an enlarged dihedral plane (Fig.~\ref{fig:AlaD-4Ch}) which clearly shows alternative and continuous paths.
The transition paths can be roughly divided into four types according to
which of the (physically equivalent) $C_{7ax}$ states they end in this extended dihedral plane. These four types of paths correspond to the combinations
of clockwise and anticlockwise rotational directions of dihedral angles $\Psi$ and $\Phi$. 
Different types of paths pass different barriers; therefore they belong to different channels.  
Under $300K$, approximately $100$ paths randomly chosen (based on their weights) are plotted in Fig.~\ref{fig:AlaD-4Ch}. 

The distributions of these four types of paths are listed in Table ~\ref{tb:4types}. Once again under $500K$, the
results given by WE method and brute-force are in good agreements. Because the enlarged plane is infinite, there are also
infinite final $C_{7ax}$ states in it. Some paths ended out of these four closest final states, the distributions of them
are listed in the ``others'' column in the table. The path ensemble connecting to the ``lower right'' $C_{7ax}$ state is the most 
important type. And when the temperature decreased, it became more dominant. From Fig.~\ref{fig:AlaD-4Ch}, we can
tell there are two channels in this type of paths. They are divided apart by the high barrier around $\Psi=0$ and $\Phi=0$.

\section{Conclusions}\label{Sec:con}
Our results show that the weighted ensemble (WE) path sampling approach is naturally suited to the fundamental problem of sampling multiple reaction channels in molecular systems.
The ability to sample multiple channels is more than an abstract statistical mechanics issue:
after all, it would not seem possible to find an ``optimal'' path without the ability to fully traverse path space.

Our studies employed toy systems to illustrate the basic mechanims underlying WE and the more familiar transition path sampling (TPS) methods, and then focused on a molecular system in detail.
In the toy systems, the differences between WE and TPS were clear:
because TPS is a Monte Carlo simulation, it can be trapped in one path channel.
By contrast, the WE method is based on the unbiased replication and combination and of brute-force simulations, and cannot be trapped.
The WE approach was also employed to find transition events between the $C_{7eq}$ and 
$C_{7ax}$ states of atomistic alanine dipeptide at temperature $500K$. The results were checked and supported by brute-force simulations.
Finally, the WE method was employed to study the same transition at room temperature $300K$. 
A high-quality transition path ensemble including multiple, clearly separated channels was obtained.

We do not know of any previous report of a statistically rigorous path ensemble  of alanine dipeptide.
Put another way, this appears to be the first report of the fractional importance of the various channels to the path ensemble, despite the small size of the molecule.

In the long term, we belive WE path sampling has key strengths, and also some weaknesses.
WE is easy to implement and suitable for use with almost type of stochastic dynamics --- including MD with a stochastic thermostat.
~\cite{Zuckerman_Jasnow-Arxiv}
The multiple-trajectory ``architecture'' of WE makes it straightforward to parallelize.
WE does not require precise knowledge of a reaction coordinate
~\cite{Zhang_Jasnow_Zuckerman-WE_CaM-PNAS-2007,
Zuckerman_Jasnow-Arxiv}
--- indeed, no ``targeting'' was used in the alanine dipeptide simulations reported here.
Further, WE sampling simultaneously provides the reaction rate and path ensemble.
Like TPS, however, WE produces correlated trajectories in the path ensembles generated --- although this does not prevent the method from achieving efficiency.
~\cite{Zhang_Jasnow_Zuckerman-WE_CaM-PNAS-2007}
Ultimately, the great flexibility of the approach, especially with regard to binning strategies 
~\cite{Zhang_Jasnow_Zuckerman-WE_CaM-PNAS-2007,
Zuckerman_Jasnow-Arxiv}
suggests we are only at the beginning of appreciating the possibilities of Huber and Kim's seminal strategy.
~\cite{Huber_Kim-WEB-BJ-1996}

%

\newpage

\newpage

\begin{table}[H]
\begin{center}
\begin{tabular}{|c|c|c|c|}\hline 
$\alpha_L$      &$C_{7eq}$     &$\alpha_R$     &$C_{7ax}$\\
\hline 
(55.1,46.4)     &(-77.9,138.4) &(-75.6,-39.9)   &(61.4,-71.4)\\
\hline 
\end{tabular}
\caption{Four energy minima of alanine dipeptide.}\label{tb:aladmin}
\end{center}
\end{table}

\begin{table}[H]
\begin{center}
\begin{tabular}{|c|c|c|c|c|c|}\hline 
&upper left&upper right&lower left&lower right&others\\
\hline
BF 500K&4.2\%&25.6\%&5.3\%&64.1\%&0.8\%\\
\hline
WE 500K&6.0\%&21.4\%&4.9\%&67.4\%&0.4\%\\
\hline
\hline
WE 300K&0.6\%&12.3\%&0.9\%&86.2\%&0.0\%\\
\hline
\end{tabular}
\caption{Distributions of the four principal channels in alanine dipeptide. The position of the final $C_{7ax}$ state in the extended plane of Fig.\ \ref{fig:AlaD-4Ch} was used to identify
the channnel. The fraction of paths which did not end in one of these four closest final states are listed in the 
``others'' column.}\label{tb:4types}
\end{center}
\end{table}

\end{document}